\documentclass[aps,prl,showpacs,final,twocolumn]{revtex4}

\usepackage{amsfonts}
\usepackage{amssymb}
\usepackage{amsmath}
\usepackage{amsthm}
\usepackage{mathrsfs}
\usepackage{bm}
\usepackage{graphicx}
\usepackage{textpos}

\begin{document}

\begin{textblock*}{5cm}(14cm,-0.8cm)
\framebox{\footnotesize MIT-CTP-4487}
\end{textblock*}

\title{Realization of a Josephson Switch}

\author{Zhaoxi Xiong}
\email{xiong@mit.edu}
\affiliation{Department of Physics, Massachusetts Institute of Technology, Cambridge, Massachusetts 02139, USA}
\date{\today}

\begin{abstract}

It has been proposed that intermittent weak links between spontaneously broken symmetry systems can give rise to a novel type of time crystal. Here a practical construction is analyzed in greater detail. For a film-geometry Josephson junction, an applied magnetic field in the junction region can effectively switch on and off the supercurrent. By tuning experimental parameters, one can possibly let the free energy be arbitrarily close to the equilibrium without compromising the life time. A parallel construction for superfluid $^4$He is also proposed.

\end{abstract}


\maketitle

The spontaneous breaking of time-translation symmetry, whose possibility was first speculated in Refs.~\cite{classicalTC,quantumTC}, may lead to broad applications from precise timekeeping to quantum simulations \cite{FeynmanClock}. In analogy with ordinary crystals, which have broken spatial translation symmetry, systems that display spontaneous temporal orders are named time crystals. While experimental efforts have been inaugurated \cite{Li} following the original proposal, no victory has yet been declared. On the other hand, it is noted in Ref.~\cite{intermittentWeakLink} that a somewhat different time crystal, one based on intermittent weak link between broken symmetry systems, specifically a Josephson junction, may plausibly exist. In that construct, the physical manifestation, namely the current, is separated from the conceptual time dependence of the relative phase. Here I shall present a closer analysis of the viability of such a plan, and propose a specific implementation of effectively intermittent Josephson link, or more vividly, a Josephson switch.

Previous studies \cite{classicalTC,quantumTC,Li,branched,Bruno} have been centered on near-ground state time crystal behaviors. In this work, the definition of time crystals will be generalized to allow for finite temperatures below the superconducting transition point, and proximity of the energy to the ground state value will be replaced by that of the free energy to the thermal equilibrium value. These more relaxed specifications will make time crystals more accessible experimentally, and reduce to the original setting in the zero-temperature limit.

\begin{figure}[!b]
\centering
\includegraphics[width=1.9in]{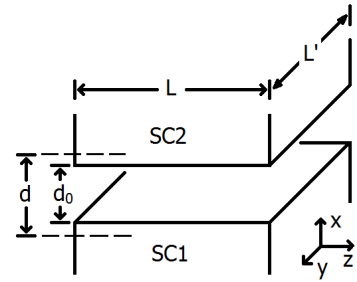}
\caption{The geometry, the coordinate system, and the dimensions of the Josephson junction. $d$ is the effective thickness.}
\label{fig:geometry}
\end{figure}

\begin{figure}[!t]
\centering
\includegraphics[width=2.1in]{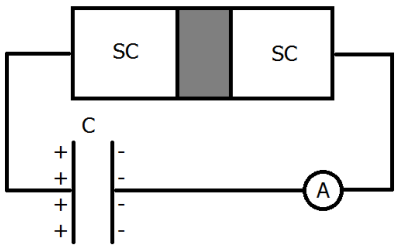}
\caption{A closed system in which a dc voltage is applied to the junction by a capacitor. The ammeter could be any other current probe. There is a uniform magnetic field in the junction region (see text).}
\label{fig:circuit}
\end{figure}

Let us consider a Josephson junction with the geometry shown in Fig.~\ref{fig:geometry}. The Josephson relations are \cite{Tinkham}
\begin{eqnarray}
j &=& \eta g (\delta) + j_{e} \left( V+\frac{1}{c}\int_{-d_0/2}^{d_0/2}\frac{\partial A_x}{\partial t}dx \right), \label{current}\\
\frac{\partial \delta}{\partial t} &=& \frac{2 e}{\hbar} \left( V+\frac{1}{c}\int_{-d_0/2}^{d_0/2}\frac{\partial A_x}{\partial t}dx \right),  \label{deltatime}\\
\nabla^{(2)} \delta &=& \frac{2 e d}{\hbar c} \bf H \times \hat x, \label{deltaspace}
\end{eqnarray}
where $j$ is the total current density in the $x$ direction, $j_e$ is the potential-dependent contribution due to quasiparticle tunneling, $\eta$ is the pair tunneling strength, $g$ is a dimensionless $2\pi$-periodic function, $\delta$ is the phase of superconductor 2 relative to superconductor 1, $V$ is the electrostatic potential of superconductor 2 relative to superconductor 1, $A_x$ is the $x$ component of the vector potential, $d_0$ is the thickness, $d$ is the effective thickness ($d_0$ plus the penetration depths on both sides), $\bf H$ is the total magnetic field in the $yz$ plane, $\nabla^{(2)}$ is the gradient operator in the $yz$ plane, and $\bf\hat x$ is the unit vector in the $x$ direction. A minimal {\it closed} system that will serve as our time crystal is shown in Fig.~\ref{fig:circuit}, where the dc voltage drop on $C$ generates a linear dependence of $\delta$ on $t$. Since net current would induce considerable dissipation in the circuit, the junction is supposed, following Ref.~\cite{intermittentWeakLink}, to permit virtually zero total current at all but designated, brief ``measurement" times. The system is immersed in a thermal bath of temperature $T$.

Conceivably, the dynamical switching of the Josephson junction can be implemented, most conveniently in point contact geometries, by separating and recombining the superconductors, but an important practical issue arises, even at zero temperature and zero bias, from the diffusion of $\delta$ while the superconductors are separated. Approximating $g(\delta) = \sin \delta$, the junction Hamiltonian can be written
\begin{equation}
H_J = E_J (1-\cos \delta) + \frac{(2e)^2}{2C_J}\hat N^2, \label{junctionHamiltonian}
\end{equation}
where $E_J \equiv \frac{\hbar}{2e}\eta \mathscr A$ is the maximum Josephson coupling energy, $\mathscr A \equiv LL'$ is the area, $C_J$ is the capacitance of the junction, and $2e \hat N $ is the charge thereon. Since $[\delta, \hat N] = i$, this Hamiltonian effectively describes the quantum mechanics of a particle of mass $\frac{\hbar^2 C_J}{4e^2}$ residing on a ring. For sufficiently large $E_J$, the ground state wave function is well-localized, and the relative phase is well-defined. Now, if $E_J$ suddenly drops to zero, as would happen when the superconductors get separated, then the wave function will spread (in the limit $C_J\rightarrow \infty$, the spreading will not happen). As the width reaches $2\pi$, a definite $\delta$ will no longer be identifiable. The time scale associated to this process is $10^{-8}$s for typical junction parameters (note that separating the superconductors further lowers $C_J$) \cite{Leggett,Sols}, so in order to have a time crystal at all, one must maintain a nonzero $E_J$, that is, a physically unbroken link.

The twin goals of having a physically unbroken link and allowing virtually zero total current can be achieved by applying a magnetic field, which henceforth will be assumed to be in the $y$ direction. In the limit $\lambda_J \gg L$, where $\lambda_J \equiv \sqrt\frac{\hbar c^2}{8 \pi e d \eta}$ is the Josephson penetration distance, self-field effects are negligible, $\bf H$ is nearly uniform, and $\delta$ depends linearly on $z$:
\begin{equation}
\delta = -\frac{2 e d}{\hbar c} H_y z + \delta(z=0), \label{phasedistribution}
\end{equation}
whence one derives the total supercurrent
\begin{equation}
I_{s} = \eta \mathscr A \frac{\sin(\pi \Phi/\Phi_0)}{\pi \Phi/\Phi_0} \sin \left[ \delta \left(z=0\right) \right], \label{fraunhofercurrent}
\end{equation}
where $\Phi = H_y L d$ is the flux through the junction, $\Phi_0 \equiv \frac{h c}{2e}$ is the flux quantum, $z=0 $ represents the center, and $g(\delta)$ has been approximated by $ \sin \delta$. One thus sees that the total supercurrent is externally switched off for nonzero integer values of $\frac{\Phi}{\Phi_0}$. Moreover, when $\Phi$ and $\Phi_0$ are of the same order, the relative phase $\delta$ only varies slowly with $z$, so the junction free energy still locally outweighs the capacitive term (see Eq.~(\ref{junctionHamiltonian})) and $\delta$ is everywhere well-defined. The threshold switch-off magnetic field can be written
\begin{equation}
H^{\text{switch}} = \frac{hc}{2eLd}. \label{switchfield}
\end{equation}
It should be noted that the existence of nodes in $I_s$ does not depend on the assumption $g(\delta) = \sin \delta$. Omitting the capacitive term, the general expression for the junction free energy per unit area is $\frac{\hbar\eta}{2e} \int_0^\delta g(\delta') d\delta'$ \cite{Josephson}. Since $\delta$ is an angular variable, one must have $\int_0^{2\pi} g(\delta) d\delta = 0$, and hence $I_s = 0$ for nonzero integer values of $\frac{\Phi}{\Phi_0}$, provided that $\delta$ is linear in $z$, i.e. $\bf H$ is uniform. It is also evident that, when a dc voltage is applied and $\frac{\Phi}{\Phi_0}$ is a nonzero integer, both the total free energy and the fact $I_s = 0$ are time-independent, showing consistency with the principles of thermodynamics.

The experiment will be so conducted that the magnetic field is kept at (integer multiples of) $H^{\text{switch}}$ at all but designated ``measurement" times, during which it deviates by some amount appropriate to the probe sensitivity. Since the magnetic field is dynamical, corrections to the linear evolution of $\delta$ can in general be introduced. Specifically, due to the second term in the right-hand side of Eq.~(\ref{deltatime}), the phase change $\delta(t_f,z) - \delta(t_i,z)$ can be modified by a measurement that occurs between $t_i$ and $t_f$. I argue, however, that this complication can be avoided by arranging for the measurement to be brief, the deviation of $\bf H$ small, and $j_e$ negligible. Assuming these conditions hold, one can linearize Eqs.~(\ref{current})-(\ref{deltaspace}) around the moment of the measurement and consider the effects of $\frac{1}{c}\frac{\partial A_x}{\partial t}$ alone. The system is now invariant under time reversal: $\delta(t) \rightarrow \delta(-t),~ j(t) \rightarrow j (-t),~ A_x(t) \rightarrow A_x(-t),~ {\bf H}(t) \rightarrow {\bf H}(-t),~ {\bf E}(t)\rightarrow -{\bf E}(-t),~ \rho(t)\rightarrow -\rho(-t)$, where ${\bf E}$ is the electric field, $\rho$ is the charge density, and $\delta$ and $j$ are taken in reference to their configurations at $t=0$, around which the equations are linearized. If ${\bf H} = H^{\text{switch}}{\bf\hat y}$ is reinstated at the end of the measurement in precisely the opposite manner to how ${\bf H}$ deviates from $H^{\text{switch}}{\bf\hat y}$ at the beginning, then, the induced electric field will add up to $\frac 1c\int \frac{\partial A_x}{\partial t} dt = 0$, and $\delta$ will resume the original configuration after ${\bf H} = H^{\text{switch}}{\bf\hat y}$ is reinstated.

Thus far we have neglected the $j_e$ term in Eq.~(\ref{current}). It must be suppressed in order for the time crystal to be long-lived. For small voltage bias, one has the estimate \cite{Tinkham}
\begin{eqnarray}
I_e \equiv j_e \mathscr A &=& - \frac{1}{eR_{NN}} \int_{-\infty}^{\infty} \frac{ N(E) N(E+eV) }{ N(0)^2 } \nonumber\\
&\times& \left[  f(E) - f(E+eV) \right] dE, \label{quasiparticletunneling}
\end{eqnarray}
where $N(E)$ is the BCS density of states, $N(0)$ is the normal density of states at the Fermi level, $f(E)$ is the Fermi occupation factor, and $R_{NN}$ is the normal resistance. Due to the existence of the superconducting gap $2\Delta(T)$, $I_e$ essentially behaves like \cite{Stephen}
\begin{eqnarray}
I_e \sim - \frac{V}{R_{NN}} e^{- \Delta(T)/k_B T}. \label{quasiparticletunnelingapproximation}
\end{eqnarray}
We consider this together with \cite{Ambegaokar}
\begin{equation}
E_J \equiv \frac{\hbar \eta \mathscr A}{2 e} = \frac{\pi \hbar \Delta(T)}{4 e^2 R_{NN}} \tanh \left(\frac{\Delta(T)}{2 k_B T}\right), \label{etaexpression}
\end{equation}
which ought to outweigh $\frac{(2e)^2}{2C_J}$ so that $\delta$ is well-defined. While it appears that $I_e$ can be suppressed and $E_J$ can be enhanced by lowering the temperature $T$, one must ensure that the coherence length $\xi(T)$ is still greater than the junction size \cite{Helium4Th} -- the lowest permissible $T$ corresponds to $\xi \sim \max\{L,L'\}$. Assuming $T \lessapprox T_c$, with $T_c$ the transition temperature, the Ginzburg-Landau theory gives \cite{GinzburgLandau}
\begin{equation}
\xi(T) = \xi_0' \left( 1- \frac{T}{T_c} \right)^{-\frac{1}{2}}, \label{coherencelength}
\end{equation}
where $\xi_0'$ is a material-dependent constant. On the other hand, the BCS theory gives \cite{BCS}
\begin{eqnarray}
\frac{\Delta(T)}{k_B T_c} &\approx& 3 \left( 1- \frac{T}{T_c} \right)^{\frac12}. \label{gapTdependence}
\end{eqnarray}
At $\xi(T) \sim \max \{L, L' \}$, one thus has
\begin{eqnarray}
I_e &\sim& - \frac{V}{R_{NN}} \exp\left[ - \frac{3\xi_0'}{\max\{L,L'\}} \right], \label{Iebest}\\
\frac{E_JC_J}{2e^2} &\sim&  \frac{C_J}{R_{NN}} \frac{5 \pi \hbar k_B T_c}{8 e^4} \left( \frac{\xi_0'}{\max\{L,L'\}} \right)^2.
\end{eqnarray}
Since $R_{NN} \propto \mathscr A^{-1}$ and $C_J \propto \mathscr A$, one finds both $I_e$ and $\frac{E_JC_J}{2e^2}$ decrease with ever decreasing $\mathscr A$, and that $\frac{E_JC_J}{2e^2}$ is maximized for a given $\mathscr A$ at $L=L'$. Presuming $\frac{E_JC_J}{2e^2} \gg 1$ is the less stringent requirement, relatively small square junctions are then preferred to large rectangular ones. This is compatible with the condition $\lambda_J \gg L$ for the uniformity of $\bf H$ and has the coalesced merit of low radiation. Also note that large $\xi_0'$ is in a transparent way desirable, and that S-c-S junctions have in general lower dissipation than S-N-S junctions.

To reduce the dissipation in generating the magnetic field, one can use superconducting rings to supply the switch-off field and a coil to generate a small additional field during measurements. Alternatively, strain can be applied to displace the superconducting rings.

One should perceive the measurement events discussed above as being external. In principle, the time crystal could measure the time spent by a specific process such as the execution of a series of logical operations, and in this respect, it is a virtually non-dissipative timekeeping device. The measured time is modulo $\frac{\pi \hbar}{e V}$, which is often inconveniently short, but some beating arrangement might help to circumvent this difficulty.

Whether measurements are to be performed or not, the time translation symmetry is measurably broken. For a given $V$, that is a given frequency, the free energy of the whole system, measured from the equilibrium, is proportional to $C$. On the other hand, Eq.~(\ref{Iebest}) indicates that the life time of the linear evolution of $\delta$ is proportional to $C LL' \exp\left[ - \frac{3\xi_0'}{\max\{L,L'\}} \right]$. By tuning $C$, $L$, and $L'$, one can let the free energy be, basically, arbitrarily close to the equilibrium without compromising the life time. A different take on this symmetry breaking matter would be to regard $V$ as a ``boundary condition." The system is then at strict equilibrium, and the suppression of $j_e$ renders the boundary condition time-independent. From either angle we see some characteristics of {\it spontaneous} symmetry breaking. In certain applications one might also be concerned with thermal noises, which would bring in the rough temporal analog of crystalline defects; we here simply point to Ref.~\cite{Stephen} where relevant issues are investigated.

\begin{figure}[!b]
\centering
\includegraphics[width=3.1in]{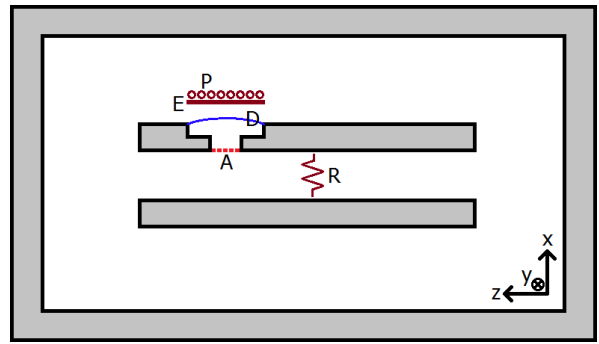}
\caption{The design for the intermittent superfluid $^4$He Josephson junction. A long, narrow flow channel, with a heat source (R) in the middle, is formed in a large closed container. A two-dimensional array of apertures (A) replaces part of a channel wall, and, together with a diaphragm (D), defines the boundary between two $^4$He reservoirs. An applied voltage between the diaphragm and a nearby electrode (E) generates a pressure on the diaphragm, which is transferred to the fluid. The displacement of the diaphragm is probed by the pickup coil (P) of a dc superconducting quantum interference device (SQUID) or any other position sensor.}
\label{fig:superfluidGeometry}
\end{figure}

A parallel construction for superfluid $^4$He appears possible. Inspired by Refs.~\cite{Helium4JosephsonDiscovery,Helium4crossover,Helium4gradient}, we consider the system shown in Fig.~\ref{fig:superfluidGeometry}. The two-dimensional array of apertures, the mass flow through them, and the pressure between the two reservoirs applied via the diaphragm, are the analogs of the weak link, the current, and the voltage, respectively. Eqs.~(\ref{current})(\ref{deltatime}) hold after suitable substitutions. The phase gradient $\nabla \phi$ in a superfluid $^4$He, in general, is proportional to the velocity $v_s$ of the superfluid component:
\begin{equation}
v_s = \frac{\hbar}{m_4} \nabla \phi,
\end{equation}
where $m_4$ is the mass of $^4$He. A nonzero $\nabla \phi$ could arise from either a boost of the entire superfluid or, as demonstrated in Ref.~\cite{Helium4gradient}, a heat flow, since the superfluid component will flow towards the heat source with the normal component flowing in the opposite direction carrying away the entropy. For the system in Fig.~\ref{fig:superfluidGeometry}, one has approximately \cite{Helium4gradient}
\begin{equation}
\frac{\partial \delta}{\partial z} = \frac{m_4  v_{s0}}{\hbar} - \frac{\rho_n}{\rho_s(\rho_s+\rho_n)} \frac{\dot Q}{2T s \sigma},
\end{equation}
where $\delta$ is the relative phase, $v_{s0}$ is the velocity of boost in the channel \cite{boost}, $\dot Q$ is the power of heating, $s$ is the specific entropy, $\sigma$ is the cross-sectional area of the channel, $\rho_s$ is the superfluid density, $\rho_n$ is the normal fluid density, and the factor $\frac12$ accounts for the two possible directions of entropy flow. Suppose $\dot Q = 0$, that a constant pressure is applied, and that $v_{s0}$ is such that the change of $\delta$ across the array is an integer multiple of $2\pi$. Then the superflow through different apertures will cancel each other, leading to zero total flow. If the normal flow is negligible, then there will be no work being done by the diaphragm. The constant pressure generates a linear dependence of $\delta$ on $t$, which is hidden since the total flow vanishes. During measurement, the heater will be turned on and the configuration of $\delta$ perturbed. Now that exact cancelation of superflow no longer occurs, the underlying time dependence is manifested in a physically measurable form. In contrast to superconducting Josephson junctions, a superfluid $^4$He Josephson junction has the virtue of being radiation-free.

\begin{figure}[!t]
\centering
\includegraphics[width=2.3in]{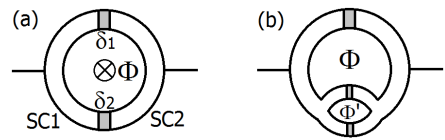}
\caption{The dc SQUID structure, originally suggested in Ref.~\cite{intermittentWeakLink} as a possible implementation of the switch. (a) The basic design, wherein the junctions are assumed to be identical. (b) A refined design wherein one of the junctions is replaced by a dc SQUID with tunable effective Josephson coupling \cite{Makhlin}.}
\label{fig:squid}
\end{figure}

I have been focused on extended junctions for which there is a field of $\delta(y,z)$ at each given time. Discretizing, one obtains a dc SQUID structure where the relative phase has only two degrees of freedom (Fig.~\ref{fig:squid}). The total current can be switched on and off by changing the threaded flux $\Phi$. The switch-off fluxes do depend on $\frac{\delta_1+\delta_2}{2}$ for general $g(\delta)$, though additional charges and currents might compensate for this in the way alluded to in Ref.~\cite{intermittentWeakLink}. I shall not discuss this alternative in detail.

In summary, I have proposed possible implementations of effectively intermittent superconducting as well as superfluid $^4$He Josephson junctions, using established technologies, as candidates for time crystals. Some characteristics of spontaneous symmetry breaking have been demonstrated. It would be intriguing to study how the constructions can extend to cold atomic \cite{BECJosephsonDiscovery,BECacdcJosephson,BECJosephsonArray} or magnon \cite{spinSFmagnonBEC} Bose-Einstein condensates.

I wish to thank Frank Wilczek, Alfred Shapere, Chetan Nayak, Zhexuan Gong, and Wujie Huang for their valuable comments.


\begin{thebibliography}{99}
\bibitem{boost} A nonzero $v_{s0}$ might be set up by cooling properly shaped, slowly rotating container through the superfluid transition \cite{rotatingHelium}, by using ``stirrers" analogous to those employed to create vortices in trapped Bose-Einstein condensates \cite{stirredBEC}, or perhaps by inserting partitions with slits initially narrow enough for phase slippage to occur while the superfluid is driven by some auxiliary diaphragm \cite{gyroscope}.
\end{thebibliography}
\end{document}